# Dependence of geosynchronous relativistic electron enhancements on geomagnetic parameters

A.V. Dmitriev[1,2] and J.-K. Chao[1]

[1]Institute of Space Science National Central University, Chung-Li, 320, Taiwan
[2]Scobeltsyn Institute of Nuclear Physics, Moscow State University, Moscow, 119899, Russia

Short title: GEOSYNCHRONOUS RELATIVISTIC ELECTRON ENHANCEMENTS

**Abstract.** Relativistic electron fluxes observed in geosynchronous orbit by GOES-8 in 1997 to 2000 were considered as a complex function of geomagnetic indices *PC, Kp,* and *Dst* as well as parameters of the magnetosphere size, subsolar *Rs* and terminator *Rf* magnetopause distances. A geosynchronous relativistic electron enhancement (GREE) is determined as daily maximal electron flux exceeding the upper root mean square deviation (RMSD) threshold of about 1500 (cm$^2$s sr)$^{-1}$. Comparison analysis of the GREE dynamics and geomagnetic conditions on the rising phase of current solar cycle revealed suppression of the relativistic electron enhancements by substantially increased strong geomagnetic activity in the solar maximum. Statistical consideration of a relationship between the GREEs and the geomagnetic parameters showed that the most important parameters controlling the geosynchronous relativistic electron enhancements were 4-day averaged *Kp* index, *PC* index and magnetopause termination distance *Rf* delayed respectively on 3 and 14 hours. Relatively high averaging time for *Kp* was explained by cumulative effect of substorm energy release in a gradual mechanism accelerating the relativistic electrons in the magnetosphere. Very short time delay for *PC* index was interpreted as intensification of a fast acceleration mechanism producing the GREEs during severe geomagnetic storms. Substantial increase of the *PC* index (*PC>5*) was founded as a sufficient condition for GREE occurrence. The fast response of the geosynchronous relativistic electron fluxes on the magnetosphere compression was explained by drift losses of the energetic electrons at the magnetopause which approaches the Earth during geomagnetic storms.

## 1. Introduction

Close relationship of the geosynchronous relativistic electron enhancements (GREE) with geomagnetic activity was founded already in the first observations of the outer radiation belt (ORB). *Rothwell and McIlwain* [1960] formulated a general pattern of the relativistic electron dynamics during a geomagnetic storm. In the beginning of geomagnetic storm the trapped relativistic electron fluxes decrease significantly while the electron precipitation increases indicating to strong losses of the trapped relativistic electrons from the ORB. On the recovery phase the intensity and average electron energy increase and exceed their prestorm levels due to either injection of energetic electrons from the interplanetary medium into the trapping zone or some acceleration process connected with the geomagnetic disturbance. This phenomenological pattern was confirmed later by numerous observations of relativistic electron fluxes in the heart of the ORB [*Arnoldy et al.*, 1960; *Hoffman et al.*, 1962; *Forbush et al.,* 1962]. Direct comparison of the electron fluxes in the ORB and in the interplanetary medium [*Van Allen and Lin*, 1960] shown that the ORB was developed over 2-day period after the storm maximum and it contained several thousand times the intensity of electrons that was present in the original solar plasma cloud before its arrival at the Earth. These facts provide convincing evidence in favor of acceleration of the relativistic electrons inside the magnetosphere.

*Paulikas and Blake* [1976] analyzing relativistic electron fluxes in geosynchronous orbit propose the following qualitative explanation of their dynamics. The changes in the electron fluxes are associated with major disruption of the energetic electron population by magnetic storms as seems in *Dst*-variation. The magnetospheric substorms are considered as the basic process which energizes magnetospheric plasma and transports accelerated particles into the stable-trapping region of the magnetosphere. This process produces the geosynchronous relativistic electron enhancements (GREEs) on the recovery phase of magnetic storms accompanied by strong substorm activity which is characterized by geomagnetic *Kp* index.

An important property of 27 day recurring periodicity of the relativistic electron enhancements in the ORB [*Williams and Smith*, 1965; *Williams*, 1966; *Paulikas and Blake, 1979; Baker et al.*, 1997] helps substantially in understanding the physics of their origin. It was revealed that the relativistic electron enhancements were associated mostly with so called recurrent magnetic storms (RMS) caused by corotating interplanetary sector structure pass the Earth [*Wilcox and Ness,* 1965; *Tsurutani et al.,* 1995]. The RMS is accompanied by intensive long-lasting (up to 10 days and more) substorm activity caused by fast and strong fluctuations of the interplanetary magnetic field (IMF).

Energetic electron injections is the most typical phenomenon observed at geosynchronous orbit during the substorms [*Parks and Winkler,* 1968]. Different models such as a substorm dipolarization mechanism [*Liu and Rostoker*, 1995] and a convection-diffusion model [*Fok et al.,* 2001], were proposed to explain the electron acceleration in the substorms. However observations [*Lezniak and Winckler*, 1970; *Kim et. al.,* 2000] show that during the substorms the electrons are effectively accelerated up to only few hundreds of keV and an ordinary substorm is able to produces only a few percents of the number of MeV electrons observed in a typical poststorm outer belt electron enhancement. Soft electrons ($E_e$<300 keV) have soft spectra with dynamics related very closely to *Kp* index in opposite to hard spectra of the relativistic electrons [*Freeman,* 1964; *West et al.,* 1981; *Cayton et al.,* 1989]. The difference in the spectral characteristics and substorm-associated variations proofs the difference in mechanisms that accelerate soft and relativistic electrons in the magnetosphere. Recently, *Ingraham et al.* [2001] showed that extremely intensive substorms during great magnetic storm are able to effectively accelerate the electrons up to relativistic energies providing the GREE. However the great magnetic storms are occasional while the GREEs are observed more often and are associated usually with moderate or severe RMS.

Hence, the substorms themselves generate mostly initial population of the soft electrons that should be accelerated up to relativistic energies by another magnetospheric mechanism. Two basic magnetospheric processes are considered for acceleration of the relativistic electrons in the magnetosphere: radial diffusion and wave-particle interaction with ULF pulsations. Direct observations of the radial diffusion in the ORB [e.g. *Frank et al.*, 1964; *Lanzerotti et al.,* 1970; *West et al.*, 1981; *Li et al.,* 1997] show convincingly that on the recovery phase of geomagnetic storms, the electrons drift across L-shells from the outer magnetosphere toward the Earth. A recirculation model [*Nishida,* 1976; *Baker et al.*, 1986; 1989] was developed to energize the soft electrons up to a few MeV during their radial diffusion from the outer to the inner magnetosphere. Comprehensive analysis of experimental results [*Schulz and Lanzerotti,* 1974; *Baker et al.*, 1989] reveals a problem with the radial diffusion timescale because it takes much too long for an electron to diffuse to low *L* values. However the radial diffusion in the ORB can increase significantly (tens times) in response to increasing geomagnetic activity represented in *Kp* index [*Lanzerotti and Morgan*, 1973; *Lanzerotti et al.,* 1978].

Global ULF waves are considered as an energy source of fast acceleration of the soft electrons up to relativistic energies in a drift resonance mechanism [*Elkington et al.,* 1999] and in an internal magnetic pumping mechanism [*Liu et al.,* 1999]. The ULF wave power increases in the beginning of the RMS when the ULF pulsations are generated by the Kelvin-Helmholtz instability which is occurred in interaction of the fast solar wind streams with the magnetopause [e.g. *Engebretson et al.,* 1998 and references therein]. In the drift resonance mechanism a convection electric field is required to increase an energization rate. On the other hand the convection electric field causes the ring current formation which leads to adiabatic betatron decceleration of trapped particles at *L*<8 and to decrease of the particle fluxes on the main phase of geomagnetic storms [*Dessler and Karplus*, 1961]. On the recovery phase the ring current decays and the particle fluxes should restore up to prestorm levels [e.g. *McIlwain*, 1996]. Experimental studies of the betatron effect on the main phase of geomagnetic storms [*McAdams and*

*Reeves,* 2001] reveal a large contribution to the relativistic electron dynamics from non-adiabatic processes. *Green and Kivelson* [2001] apply the drift resonance mechanism as well as the betatron effect for prediction of the relativistic electron fluxes during several geomagnetic storms. They obtain that one of nine relativistic electron enhancement events can not be explained by the drift resonance model and conclude that simply having ULF wave power at the resonant frequency is not alone sufficient to cause a relativistic electron enhancement.

The internal magnetic pumping mechanism proposed by *Liu et al.* [1999] requires intensive scattering of the accelerating electrons to violate the third adiabatic invariant and, thus, to avoid trivial operation of the betatron mechanism. Recently, *Kanekal et al.* [2001] obtain remarkable coherence of the relativistic electron dynamics through the ORB indicating to fast isotropization timescales for these electrons (of the order of half day at most) which is associated with effective pitch-angle scattering of the electrons. On the other hand the scattering leads to intensive electron precipitations and losses that is usually observed at $L$=4-6 during the GREEs [*Imhof et al.,* 1991]. *O'Brien et al.* [1964] find that the precipitation of relativistic electrons increases exponentially up to ten times in response to increasing $Kp$ index. Such a dependence of the relativistic electron losses on geomagnetic activity imposes strong requirements on the effectiveness of accelerating mechanisms.

Due to the intensive losses a lifetime of the relativistic electrons generated in the fast mechanisms on the main phase of geomagnetic storms is restricted by 3 to 4 days [*Baker et al.,* 1986]. *Dmitriev et al.* [2001] suggest the mechanism of magnetic pumping on the tail magnetopause to produce a long-duration acceleration of the relativistic electrons in numerous interactions with ULF magnetic field fluctuations on the magnetopause of the Earth's magnetotail. Such fluctuations can be surface waves generated by the Kelvin-Helmholtz instability or by substorm activity. However even during intensive geomagnetic activity the fluxes of energetic electrons in the magnetosheath layer near the magnetopause are not sufficient to produce fluxes in GREEs [*Baker and Stone,* 1977], therefore accumulation of the relativistic electrons in the outer magnetosphere is required.

The survey of the existing accelerating mechanisms gives ambiguous picture of a relationship of the GREE with parameters of geomagnetic activity. Undoubtedly, substorm activity (characterized by *AE,* *Kp* or *Ap* indices) produces enhancements of high energy (up to hundreds of keV) electrons in the outer magnetosphere but subsequent relativistic electron acceleration depends on different factors sometime in opposite manner. The ring current formation (characterized by the *Dst* variation) leads to decrease of the trapped electron fluxes due to the betatron effect. On the other hand, the convection electric field responsible for the ring current intensification increases the rate of the electron energization in the drift resonance mechanism. The *Kp* index increase promotes to effective operation of the recirculation and magnetic pumping mechanisms. At the same time significant increase of the *Kp* leads to intensive electron losses in the outer magnetosphere.

In this situation empirical models become very popular. Linear filters were developed for prediction of the geosynchronous relativistic electron fluxes from geomagnetic *Kp* [*Nagai*, 1988] and *AE* indices, and solar wind velocity [*Baker et al.,* 1990]. These filters describe reasonably only average variations of the electron fluxes with 2-3 days delay for the daily sum *Kp* index and 1 day delay for daily mean solar wind speed. *Koons and Gorney* [1991] successfully apply an artificial neural network (ANN) technique for day-ahead forecasting of the geosynchronous relativistic electron fluxes from 10-day input set of daily sum *Kp*. *Freeman et al.* [1998] develop an ANN model of the geosynchronous relativistic electron fluxes on the main phase of geomagnetic storms using *Dst* variation and low energy electron fluxes as input parameters of the network. Recently, *Li et al.* [2001] propose advanced nonlinear filter to predict geosynchronous relativistic electron fluxes using *Dst* index and solar wind parameters. Electrons are accelerated by radial diffusion from the magnetopause with the coefficient depending on the solar wind parameters. The model also includes betatron effect and effect of the electron losses due to magnetosphere compression by the solar wind dynamic pressure.

Different theoretical and empirical models represent geosynchronous relativistic electron fluxes as a function of various geomagnetic parameters. A nature of increases and decreases of the relativistic electron fluxes in geosynchronous orbit can be different. On the other hand space weather requires in the first turn a careful prediction of strong enhancements of the geosynchronous relativistic electrons that cause dangerous internal charge in the onboard electronic equipment of satellites. Hence studies of the GREEs are useful and important for understanding of their drivers and acceleration processes. In following section of the present study we determinate the geosynchronous relativistic electron enhancements from consideration of the GOES-8 data in 1997 to 2000. The GREE and geomagnetic activity variations on the rising phase of the current solar cycle are analyzed in the third section. Correlation analysis and multi-parametric linear regression between the GREEs and geomagnetic parameters are performed in the fourth section. The fifth section is devoted to investigation of necessary and sufficient geomagnetic conditions for the GREE occurrence. Discussion and conclusions are presented in the sixth section.

## 2. Initial data

High time resolution (1.5 min) data on relativistic electron fluxes in geosynchronous orbit are obtained from the ISTP database (ftp://cdaweb.gsfc.nasa.gov/pub/istp/) of GOES-8 measurements of >2 MeV electrons in 1997 to 2000. We study maximal values of the relativistic electron fluxes observed in a range of local time LT=12±6h. Such choice permits to sense fast (hours) dynamics of the relativistic electrons. Undoubtedly the orbital satellite rotation influences on observing temporal dynamics of the electron fluxes. In according to estimation by *Paulikas et al.* [1968] omnidirectional relativistic electron flux in this local time range is produced by the electrons arriving from space range of $L{\approx}6.1{\sim}6.9$ at noon to $L{\approx}6.3{\sim}7$ at flanks. In the fist approach we can neglect such local time dependence for the electron fluxes. Unfortunately only GOES-8 data covers the entire time interval of 1997 to 2000. Involving in the analysis GOES-9 and GOES-10 measurements at other local time requires additional procedure of the satellite inter-calibration. This complicated subject is beyond the frame of the present work.

Figure 1 presents time profile of the daily maximal geosynchronous electron flux in 1997-2000 together with *Kp*, *Dst* and *PC* geomagnetic indices as well as hourly minimal subsolar *Rs* and termination *Rf* magnetopause distances calculated in Earth's radii (Re). Information about *Kp*, *Ap* and *Dst* geomagnetic indices is obtained from NGDC public data service (ftp://ftp.ngdc.noaa.gov/). Hereafter we consider *Kp* index in decimal scale (multiplied on 10). Hourly averaged *PC* index (http://www.aari.nw.ru/clgmi/geophys/index.htm) is calculated from measurements of the polar geomagnetic variations [*Troshichev et al.*, 1988]. The distances *Rs* and *Rf* are calculated using *Chao et al.* [2002] magnetopause model and ISTP high-resolution data from WIND satellite in 1997 and from ACE satellite in 1998 to 2000. Minimal distances *Rs* and *Rf* are calculated for each hour in 1997-2000 when upstream data are available. In Figure 1 one can clearly see that maximal relativistic electron flux in geosynchronous orbit varies significantly (more than three orders of magnitude). There is no prominent trend in the electron fluxes on the rising phase of the current solar cycle. But a seasonal variation with minimum in summer (June-July) and winter (December-January) corresponding to minimum in the geomagnetic activity is prominent especially in 1997 to 1999. In the solar maximum 2000, the seasonal variation is distorted by enhanced solar wind and geomagnetic disturbances.

A distribution of occurrence number of the maximal electron fluxes in geosynchronous orbit is shown on the top panel of Figure 2. A shape of the distribution is very close to a log-normal with the most probable and average values of about 200 $(cm^2 s \ sr)^{-1}$ and the upper root mean square deviation (RMSD) from the average value of about 1500 $(cm^2s\ sr)^{-1}$. We determine the "geosynchronous relativistic electron enhancement" (GREE) as daily maximal flux of >2 MeV electrons in geosynchronous orbit with magnitude higher than upper RMSD threshold equal to $1500(cm^2s\ sr)^{-1}$. A distribution of occurrence number of local time for maximal electron flux observations is presented on the bottom panel of Figure 2 by solid histogram (left axis) for all maximal electron fluxes and by dotted histogram (right axis) for the GREEs. The most probable location both for the maximal relativistic electron fluxes and for the enhancements is placed in vicinity of the noon (LT=9~14h) with a shift toward morning hours with median LT~11.5 hours. The asymmetry of 0.5 hour in the location of the maximal electron fluxes relative to the noon can not be simply explained by well known aberration effect of the Earth orbital rotation. Indeed, if we accept the average solar wind velocity 400 km/s than aberration effect of the Earth rotating with velocity 30 km/s should be only 0.3 hour. Therefore, the asymmetry is connected with the spatial structure of the magnetic filed in the outer magnetosphere, which is controlled by both shield current on magnetopause associated with compression and inner magnetospheric currents such as ring current, tail current and field align currents associated with storm and substorm geomagnetic activity.

Detail consideration of Figure 1 permits revealing the GREEs on February 13-14, 1998; December 5-9 1998, December 5-12, 1999; February 25-29, 2000; and December 11-16, 2000 when geomagnetic activity is relatively weak: Dst>-50 nT and Kp<50. The List of Solar Proton Events Affecting the Earth Environment (http://sec.noaa.gov/ftpdir/indices/SPE.txt) does not give substantial solar particle fluxes in 5-day vicinity ahead these intervals. Hence even moderate geomagnetic storm is not necessary condition for the GREE occurring. On the other hand we can find at least 8 strong geomagnetic storms (Kp>60 and/or Dst<-100 nT) that do not produce GREEs in 5-day vicinity after the storm maximum. There are magnetic storms on April 10-11, 1997; October 1, 1997; November 7, 1997; March 21, 1998; April 16-17, 1999; September 22-23, 1999; August 12, 2000; and October 4-5, 2000. Therefore strong magnetic storm sometime is not sufficient to produce GREEs.

## 3. Relation to the solar activity

A relationship between the GREEs and geomagnetic activity is exhibited in their variations with solar activity. Here we study these variations on the rising phase of the current 23[rd] solar cycle. Figure 3 presents a number of days per year with strong and continuous substorm activity, when daily $Ap$>20 (dashed line), and with GREE (solid line). The solar activity is characterized by the annual sunspot number indicated by dotted line (right axis). The planetary geomagnetic indices $Kp$ and $Ap$ are significantly contributed by auroral electrojet and correlate very well with $AE$ index directly associated with substorm activity [*Rostoker*, 1972, 1991]. Hence hereafter we consider the indices $Kp$ and $Ap$ as indicators of the substorm geomagnetic activity. The number of days with disturbed $Ap$ index growths gradually on 3 times from 1997 to 2000. Such dynamics reflects general increase of the interplanetary medium perturbation from the solar activity minimum to the maximum due to increase of a number of interplanetary transient events. These events influence also on the GREE occurrence frequency. The number of days per year when GREEs are observed increases substantially from 43 in 1997 to 79 in 1998 and 66 in 2000. It is interesting to note that the number of days with strong substorm activity and the number of days with the GREEs are comparable. However the dynamics of their growth is different. The number of days per year with GREEs has rapid increase in the beginning of the rising phase and then gradually decreases toward the solar maximum. This fact should mean that very intensive geomagnetic activity suppresses the GREE.

To study a problem of suppression we consider GREE intervals when geosynchronous relativistic electron enhancements are lasting continuously from one to several days. There are 70 GREE intervals in 1997 to 2000. Integral spectra of the interval duration are presented in Figure 4 for each year from 1997 to 2000 by squares, crosses, triangles and circles respectively. The spectra $N(>t)$ are well approximated by an exponential low $N(>t)=a/\exp(t/t_0)$. One can see that on the rising phase of solar activity the spectrum $N(>t)$ becomes more gradual such that the character duration $t_0$ of the relativistic electron enhancements increases from $t_0=2.44$ in 1997 to $t_0=4$ in 1999. However in the solar maximum in 2000 the character duration is abruptly dropped down to $t_0=2.5$. On Figure 5 we present the number of days per year with strong geomagnetic activity when $Dst<$-100 nT (solid line), 3-hour $Kp>6.5$ (dashed line) and 1 hour $PC>10$ (dotted line). The number of GREE intervals per year is presented by thick line. Contrary to substantial increase of the strong magnetic storm number (up to four times) from few in 1997 to 15~20 in 2000, the number of GREE intervals practically does not change. Thereby the geosynchronous relativistic electron enhancements should be suppressed by strong geomagnetic disturbances resulting to decrease the GREE occurrence frequency and the duration of GREE intervals in the maximum of solar activity.

## 4. Statistical analysis

To estimate numerically a relationship between the GREEs and geomagnetic parameters we suggest that relativistic electron flux $I(t)$ in geosynchronous orbit is a function of particle sources and losses depending on geomagnetic activity. The sources $Q$ are considered as a function of the geomagnetic parameters ($GP$) including a time delay $\Delta t$ required for the electron acceleration. The electron losses are characterized by decay time $\tau$ which is also function of geomagnetic parameters. Hence we can present the equation for dynamics of the relativistic electron flux in geosynchronous orbit in following form:

$$\frac{dI_e(t)}{dt} = Q(GP(t-\Delta t)) - \frac{I_e(t)}{\tau(GP(t))}, \qquad (1)$$

The value of $\Delta t$ is estimated from several hours [*McAdams and Reeves,* 2001] to 2~3 days [*Baker et al.*, 1986]. The loss time $\tau$ varies from 3 to 4 days [*Baker et al.,* 1986] and it is comparable with character time of the GREE interval duration $t_0$ varying from ~2.4 to 4 days. It means that the electron losses during the GREE are essential and comparable with the electron sources. Indeed as one can see in Figure 2 many GREE time profiles have a "step-like" shape with fast increase in the beginning and fast decrease in the end of the intervals. During such GREE intervals the electron flux variations are relatively weak (less than order of magnitude) indicating the balance between the particle sources and losses. Therefore in the case of the GREEs, the equation (1) can be approximately converted to:

$$I_e(t) \sim \tau(GP(t)) \cdot Q(GP(t-\Delta t)) \equiv P\{GP(t), GP(t-\Delta t)\}. \quad (2)$$

Here $P\{GP\}$ is production function depending both on current and on preceding geomagnetic activity. Hence even one a geomagnetic parameter can influence on the GREEs in different time scales. As we have shown above the maximal relativistic electron fluxes have a log-normal distribution. Therefore consideration of $\log(I_e(t))$ allows in the first approaching to represent the production function $P\{GP\}$ as a linear combination of geomagnetic parameters.

The cross-correlation of the $\log(I_e(t))$ in the GREEs with geomagnetic indices shifted on various $\Delta t$ is presented in Figure 6. The values of 3-hour $Kp$ index are repeated within each 3-hour intervals. The cross-correlation with $Kp$, $Dst$, $PC$, $Rs$ and $Rf$ parameters is indicated by respectively solid, dashed, dotted black lines and gray dotted and gray dashed lines. Time delay $T_d$ is calculated from a hour of the maximal relativistic electron flux observation. The dependence of the cross-correlation coefficients from the time delay is non-uniform. Partial correlation coefficients for the geomagnetic indices achieve maximum at time delay $T_d\sim 2$ days. The best partial correlation coefficients and corresponding time delays $T_d$ as well as number of available data are presented in Table 1. It is interesting to note that correlation coefficient for the geomagnetic indices (especially for the *PC*) demonstrate few local maxima reiterative with quasi 1-day periodicity. The cross-correlation for parameters of the magnetosphere size *Rs* and *Rf* demonstrates two different regimes. A relatively high correlation ($r\sim 0.2$) is revealed in vicinity of $T_d=14$ hours and highest anti-correlation $r\sim -0.3$ is observed at $T_d=75$ hours. Such behavior indicates to two different regimes of the magnetosphere size influence on the GREEs represented by the equation (2).

To study cumulative effects of the geomagnetic activity in producing relativistic electron enhancements a correlation of the GREEs with running averaged geomagnetic parameters is considered (Figure 7). The running averaged is calculated for various time intervals $T_a$ just before the hour of the maximal electron flux observation. Dependencies of correlation coefficients from averaging time $T_a$ are gradual with wide maximum in the range of 2 to 8 days. The best correlation coefficients and corresponding averaging time $T_a$ as well as number of available data are presented in the Table 1. Again the *Kp* index demonstrate highest correlation with the GREEs ($r=0.38$ for $T_a=4$ days). We have to indicate to high inter-correlation between magnetopause parameters *Rs* and *Rf* that is originated from averaging of southward and northward orientation of the IMF though southward IMF only causes changing of the magnetopause flaring. Therefore in following consideration of cumulative effects we will use only the magnetopause subsolar distance *Rs*.

The best correlation coefficients obtained in the statistical analysis (see Table 1) are actually weak ($r\sim 0.2$-$0.4$). As we have shown in the Introduction the geomagnetic parameter influence on the GREE is complicated and indirect. In this situation the multi-parametric linear regression can be useful in determination of the best set of the parameters producing a highest common correlation $r_c$ with the GREEs and, thus, in determination of the most important parameters for the production function $P\{GP\}$. As we accepted above the relationship between logarithm of intensity of the relativistic electron enhancements *Ie* and geomagnetic parameters can be presented in linear form:

$$\log(Ie) = a_{00} + \sum a_{1i} GP\big|_{T_d} + \sum a_{2i} \langle GP \rangle_{T_a} \qquad (3)$$

Here $GP\big|_{T_d}$ is hourly value of a geomagnetic parameter shifted on $T_d$ hours, and $\langle GP \rangle_{T_a}$ is $T_a$-hour averaged value of a geomagnetic parameter. The corresponding time delays and averaging times are taken from the Table 1.

A multi-parametric linear regression allows obtaining coefficients $a_{ij}$ and correlation coefficients both for entire set of the regression parameters and for each individual parameter of the regression (partial correlation coefficient). To find the best set of the parameters producing reasonable common correlation coefficient, we can take out one or other parameter from the regression and study how this removing influences on common correlation coefficient. Steps of the selection are presented in the Table 2 where the common correlation coefficient $r_c$ and the partial correlation coefficients are presented. Crosses in the Table 2 indicate that a parameter is removed from the regression. We analyze 164 GREEs for which the data on geomagnetic parameters from the Table 1 are available. When all parameters are involved in the linear regression the common correlation coefficient is $r_c\sim 0.6$. Removing of the parameters $Kp\big|_{45}$, $Rs\big|_{75}$, and $\langle PC \rangle_{96}$ does not influence on the common regression coefficient because they have highly inter-correlation with other parameters. When we take out the parameters $Dst|_{44}$, $\langle Dst \rangle_{78}$ and $\langle Rs \rangle_{182}$ the common correlation coefficient decreases slightly on one to two of tenth that indicates on relatively small contribution of this parameters in the regression. Removing parameters $PC|_3$, $Rf|_{14}$ and $\langle Kp \rangle_{96}$ from the regression leads to substantial decrease of the common correlation coefficient. Moreover a set from these three parameters produces reasonable common correlation coefficient $r_c$=0.54.

Thereby the geomagnetic parameters $PC|_3$, $Rf|_{14}$ and $\langle Kp \rangle_{96}$ have the most significant influence on geosynchronous relativistic electron enhancements. Using values of the coefficient $a_{ii}$ from the multi-parametric linear regression we can present numerical dependence of the GREEs from the geomagnetic parameters:

$$\log(Ie) = 2.3 + 0.015 \langle Kp \rangle_{96} + 0.03 PC\big|_3 + 0.059 Rf\big|_{14} \quad (4)$$

The most important parameters in this expression are $\langle Kp \rangle_{96}$ and $Rf|_{14}$ that produce common correlation coefficient $r_c$=0.5 in the regression. The third parameter $PC|_3$ have weaker but still significant contribution in the regression providing the fast influence of the geomagnetic activity on the GREEs.

## 6. Necessary and sufficient conditions for the GREE

In the previous section we have shown that even complex inputs of linear combinations of the geomagnetic parameters in the production function $P\{GP\}$ gives only a moderate common correlation $r_c$<0.57 with the GREEs. Such a situation indicates to strong nonlinearity of the production function which can reveal itself both in a strong nonlinear shape and in a temporal properties of the $P\{GP\}$. For example, a nonlinear dependence of the production function on the $Kp$ is originated from exponential growth of the electron losses with $Kp$. The effectiveness of acceleration mechanisms increases slower with $Kp$. Therefore at some $Kp$ the rate of electron losses should exceed the rate of sources that leads to decrease of the electron fluxes with further increase of $Kp$. Obviously the correlation between the electron fluxes and $Kp$ can not be high. The other kind of nonlinearity is a dependence of the acceleration rate $\Delta t$ on geomagnetic activity $\Delta t \equiv \Delta t(GP)$. It seems too difficult to take into account all possible ways producing relationship between the GREEs and geomagnetic parameters. But we can simplify the problem by considering only extreme values of the relativistic electron fluxes and geomagnetic parameters. In this case the time derivation in the left side of the equation (1) is exactly equal to zero and we can estimate directly a relationship between extreme relativistic electron fluxes and extreme values of geomagnetic parameters which, we anticipate, contribute mostly to the production function $P\{GP\}$.

For each of 70 GREE intervals we find an extremum of the electron flux within the first 3 days of the GREE interval and corresponding maximal values of geomagnetic parameters observed during 4 days before the maximal electron flux. The left side of Figure 8 demonstrates scatter plots of maximal values of geomagnetic parameters $Kp$ (a), $Dst$ (b), and $PC$ (c) versus extreme electron fluxes. The best fit (straight lines) in semi-logarithmic scale and corresponding correlation coefficients $r$ are also presented. One can see a wide scattering of the points on the plots and relatively low correlation coefficients with maximal $r$=0.43 for $Kp$ index. Detail consideration shows that extreme electron fluxes are preceded by moderate and weak geomagnetic activity when maximal $Kp$>30 and/or maximal $Dst$>-50 nT. Moreover even extremely high relativistic electron fluxes ($I_e$>$10^5$ (cm$^2$ s sr)$^{-1}$) can be preceded by relatively low geomagnetic activity with $Kp$<50 and/or $Dst$>-50 nT. It means that strong geomagnetic activity is not necessary condition for the GREE. The same situation is observed for $PC$ index (Figure 8c). Most of the $PC$ index associated with GREEs is corresponded to slightly disturbed geomagnetic conditions in the polar region.

Estimation of the time delay $\Delta t$ between maximal relativistic electron fluxes $I_e$ and maximal values of the geomagnetic parameters reveals prominent maximum in the $\Delta t$ occurrence probability at $\Delta t$=82h for $Kp$ index, at $\Delta t$=62h for $Dst$ index and two maxima for $Pc$ index at $\Delta t$~3h and at $\Delta t$=57h. Thereby the $Pc$ index influences on the GREEs in two different time scales (fast affect and delayed influence) that indicates on different time scales of the electron acceleration in the magnetosphere.

The right panels of Figure 8 demonstrate scatter plots for the maximal electron fluxes (ordinate) determined within 4 days after daily maximal values of geomagnetic parameters (abscissa) $Kp$ (d), $Dst$ (e) and $PC$ (f) in 1997 to 2000. The spread of the points on the plots is significant especially for daily maximal $Kp$ and $Dst$ indices. In is interesting to note that *Freeman* [1964] demonstrated similar picture for relativistic electron fluxes versus $Kp$. The strong magnetic storm ($Dst$<-100) and/or substorm ($Kp$>60) activity can be followed by the GREEs as well as by average or even small electron fluxes ($I_e \leq 200$ (cm$^2$s sr)$^{-1}$). This fact indicates convincingly that strong geomagnetic storm is not sufficient condition for the GREE. Moreover as we can see on the right upper corner of the scatter plots for the $Kp$ and $Dst$ indices, after very strong geomagnetic activity the daily maximal relativistic electron fluxes have a tendency to decrease. This fact is an additional support to our conclusion about the GREE suppression by the strong geomagnetic activity. Contrary, the $PC$ index demonstrates more ranked behavior. In 20 of 24 cases (>80%) when daily maximal $PC$>5, the GREEs are observed within following 4 days. In this sense strongly disturbed $PC$ index can be considered as a sufficient condition for the GREE occurrence.

## 6. Discussion and conclusion

We find that the relativistic electron enhancements in geosynchronous orbit have close and complicated relationship with geomagnetic parameters in a wide temporal range from years to hours. The frequency of the GREE occurring increases with growth of the geomagnetic activity on the rising phase of the current 23$^{rd}$ solar cycle. However in the solar maximum the number of geosynchronous relativistic electron enhancements drops down and the GREE interval duration becomes shorter. We interpret this fact as suppression of the GREEs by strong geomagnetic activity. Indeed a long-lasting GREE associated usually with the RMS can be split by an occasional geomagnetic storm. Due to adiabatic betatron deceleration on the main phase and following acceleration on the recovery phase, the relativistic electron flux suffers substantial negative variation with period of one to two days. The occasional strong magnetic storm itself should not produce the relativistic electron enhancement. In the second section we have listed 8 strong magnetic storms without subsequent GREEs. But such a storm can split or interrupt the relativistic electron enhancements especially when it is accompanied by strong and short-duration substorm activity which is sufficient to intensify the electron scattering and losses but insufficient to provide the effective electron acceleration in the magnetosphere.

We have found that the most important geomagnetic parameters controlling the geosynchronous relativistic electron enhancements are $Kp$ index, $PC$ index and magnetopause termination distance $Rf$. These parameters operate in different time ranges that are corresponded to different time scales of the relativistic electron acceleration and losses in the magnetosphere. The ambiguity in a dependence of the GREEs on the geomagnetic activity revealed in the statistical analysis is demonstrated in Figure 9 for severe geomagnetic storm on April 30 (DOY=120) to May 5, 1998 (DOY=125). The storm sudden commencement on April 30 is owing to the magnetosphere compression by the solar wind pressure. This day the $Dst$ variation is positive and the $Kp$ index slightly increases. Due to the compression the magnetosphere size decreases such that the subsolar distance approaches close the geosynchronous orbit. The relativistic electron flux decreases abruptly on the next day (May 1, 1998). The decrease can be explained by fast escape of the soft electrons from the outer magnetosphere through approaching magnetopause to the interplanetary medium. As a result, the soft electron flux becomes insufficient to support the GREE. The relativistic electrons in geosynchronous orbit do not response immediately on the soft electron flux decrease due to two reasons: 1. The relativistic electron losses through the magnetopause are not so fast because a slow radial diffusion; 2. A certain time is required for acceleration of the electrons and/or their transport to the geosynchronous orbit. In this sense the correlation between the GREEs and the magnetopause termination distance $Rf$ delayed on 14 hours can be easily explained by influence of the magnetosphere size on the soft electron population in the outer magnetosphere. By this way the time delay of 14 hours is associated with a time of the electron acceleration and transport in the outer magnetosphere. This time scale is in a good agreement with the rate of relativistic electron enhancements estimated by *McAdams and Reeves* [2001].

On May 2 to 4, 1998 the geosynchronous relativistic electron flux has very low intensity due to three effects operating simultaneously. The magnetosphere size is relatively small such that the magnetopause crosses the geosynchronous orbit on May 4 when the weakest geosynchronous relativistic electron flux is observed. Development of the severe geomagnetic storm indicates that the magnetosphere size is decreased due to intensive erosion on the magnetopause under long-lasting southward IMF. The storm-time betatron decceleration of the electrons by developing ring current is the second effect causing relativistic electron flux decrease, especially on May 4 when the severe geomagnetic storm ($Dst \sim -200$ nT) occurs. Intensive electron scattering and precipitation in the outer magnetosphere during very strong geomagnetic activity ($Kp > 6$) is the third effect responsible for the strong losses of the trapped relativistic electrons. On May 5, 1998 the magnetosphere size increases substantially, the ring current decays and substorm activity decreases gradually. At the same time the geosynchronous relativistic electron flux increases significantly up to its prestorm value and the GREE is restored. In this event a time of the GREE development is about 1 day and, thus, the fast acceleration of the relativistic electrons takes place.

It is important to note that on the second half of May 5, the $PC$ index is still high ($PC > 10$) while the effects causing the intensive electron losses become weak. As a result the GREE beginning on May 5 is observed only a few hours after the $PC$ index enhancement. A correlation of the GREEs with 3-hour delayed $PC$ index indicates to very fast response of the relativistic electrons on geomagnetic activity in the polar region. *Troshichev et al.* [1988] show that the $PC$ index is associated with merging electric field defining the polar cap convection. At the same time the convection electric field controls the rate of fast electron acceleration in the drift resonance mechanism. *Vennerstrom et al.* [1991] found high correlation ($r \sim 0.8$) between $PC$ index and auroral electrojet $AE$ index directly connected with substorm activity. This relationship should indicate to acceleration of the relativistic electrons by the convection-diffusion mechanism. Furthermore in the previous section we have shown that strongly disturbed $PC$ index ($PC > 5$) is a sufficient condition for the GREEs. Thereby the short time delay in the cross-correlation between the GREEs and $PC$ index should be connected with effective operation of the drift resonance or convection-diffusion mechanisms producing fast relativistic electron acceleration in the magnetosphere.

The other interesting feature of the geomagnetic storm on April 30 to May 5, 1998 is post-storm continuous gradual increase of the geosynchronous relativistic electron flux on May 6-8 that indicates strictly to operation of a long-lasting acceleration mechanism in the magnetosphere. Cross-correlation analysis gives that the GREEs have the best correlation with 4-day averaged $Kp$ index. Such a long averaging time indicates to a long-lasting accumulation of the substorm energy release in the magnetosphere which produces enhancements of the relativistic electrons accelerated by a gradual mechanism including substorms as a consisting part. This fact says in favor to effective operation of the recirculation model or the mechanism of magnetic pumping on the tail magnetopause.

Finally we can conclude that GREEs are mostly produced by long-lasting (more than 4 days) substorm activity (indicated in the *Kp* index) when operation of gradual acceleration mechanisms in the magnetosphere is most effective. In severe geomagnetic storms especially with high *PC* index a fast accelerating mechanism is effective and produces fast increase of the geosynchronous electron fluxes. Strong geomagnetic activity suppresses the GREEs due to effect of betatron deceleration on the main phase of geomagnetic storms as well as electron losses due to both intensive electron precipitation during strong substorm activity and fast escape of the electrons from the outer magnetosphere through the approaching magnetopause.

**Acknowledgement** This work is supported by grants NSC-90-2811-M-008-010 and NSC-90-2111-M-232-001 from the National Science Council and grant INTAS 2000-752.

**References**
Arnoldy, R.L., R.A. Hoffman, and J.R. Winckler, Observations of the Van Allen radiation regions during August and September 1959, Part 1, *J. Geophys. Res., 65*, 1361, 1960.

Baker, D.N., and E.C. Stone, The relationship of energy flow at the magnetopause to geomagnetic activity, *Geopys. Res. Lett., 10*, 395, 1977.

Baker, D.N., J.B. Blake, R.W. Klebesadel, and P.R. Higbie, Highly relativistic electrons in the earth's magnetosphere 1. Lifetimes and temporal history 1979–1984, *J. Geophys. Res., 91*, 4265, 1986.

Baker, D.N., J.B. Blake, L.B. Callis, R.D. Belian, and T.E. Cayton, Relativistic electrons near geostationary orbit: evidence for internal magnetospheric acceleration, *Geopys. Res. Lett., 16*, 559, 1989.

Baker, D.N., R.L. McPherron, T.E. Cayton, and R.W. Klebesadel, Linear prediction filter fnalysis of relativistic electron properties at $6.6R_E$, *J. Geophys. Res., 95*, 9, 15,133, 1990.

Baker, D.N., X. Li, N. Turner, J.H. Allen, L.F. Bargatze, et al., Recurrent geomagnetic storms and relativistic electron enhancements in the outer magnetosphere: ISTP coordinated measurements, *J. Geophys. Res., 102*, 14,141, 1997.

Cayton T.E., R.D. Belian, S.P. Gary, T.A. Fritz, D.N. Baker, Energetic electron components at geosynchronous orbit, *Geopys. Res. Lett., 16*, 147, 1989.

Chao, J. K., D. J. Wu, C.-H. Lin, Y. H. Yang, X. Y. Wang, Models for the size and shape of the Earth's magnetopause and bow shock, Space Weather Study Using Multipoint Techniques, Ed. L.-H. Lyu, Pergamon, 360p., 2002.

Dessler, A.J., and R. Karplus, Some effects of diamagnetic ring current on Van Allen Radiation, *J. Geophys. Res., 66*, 2289, 1961.

Dmitriev A.V., M.F. Bakhareva, and Yu.S. Minaeva, Electron acceleration by magnetic pumping on the tail magnetopause, *Adv. Space Res., 28*, 5, 807-812, 2001.

Elkington, S.R., M.K. Hudson, and A.A. Chan, Acceleration of relativistic electrons via drift-resonsnt interaction with toroidal-mode Pc-5 ULF oscillations, *Geopys. Res. Lett., 26*, 3273, 1999.

Engebretson, M., K.-H. Glassmeier, M. Stellmacher, W.J. Hughes, and H. Luhr, The dependence of high-latitude Pc5 wave power on solar wind velocity and on the phase of high-speed solar wind streams, *J. Geophys. Res., 103*, 26,271, 1998.

Fok M.-C., T.E. Moore, and W.N. Spjeldvik, Rapid enhancement of radiation belt electron fluxes due to substorm dipolarization of the geomagnetic field, *J. Geophys. Res., 106*, 3873, 2001.

Forbush, S.E., G. Pizzelia, and D. Venkatesan, The morphology and temporal variations of the Van Allen radiation belt, October 1959 to December 1960, *J. Geophys. Res., 67*, 3651, 1962.

Frank, L.A., J.A. Van Allen, and H.K. Hills, A study of charged particles in the earth's outer radiation zone with Explorer 14, *J. Geophys. Res., 69*, 2171, 1964.

Freeman, J.W., The morphology of the electron distribution in the outer radiation zone and near the magnetospheric boundary as observed by Explorer 12, *J. Geophys. Res., 69*, 1691, 1964.

Freeman, J.W., T.P. O'Brien, A.A. Chan, and R.A. Wolf, Energetic electron at geostationary orbit during the November 3-4, 1993 storm: Spatial/temporal morphology, characterisation by a power low spectrum and, representation by an artificial neural network, *J. Geophys. Res., 103*, 26,251, 1998.

Green, J.C., and M.G. Kivelson, A tail of two theories: How the adiabatic response and ULF waves affect relativistic electrons, *J. Geophys. Res., 106*, 25,777, 2001.

Hoffman, R.A., R.L. Arnoldy, and J.R. Winckler, Observations of the Van Allen radiation regions during August and September 1959, *J. Geophys. Res., 67*, 4543, 1962.

Imhof, W.L., H.D. Voss, J. Mobilia, D.W. Datlowe, and E.E. Gaines, The precipitation of relativistic electrons near the trapping boundary, *J. Geophys. Res., 96*, 5619, 1991.

Ingraham, J.C., T.E. Cayton, R.D. Belian, R.A. Christensen, R.H. Friedel et al., Substorm injection of relativistic electrons to geosynchronous orbit during the great magnetic storm of March 24, 1991, *J. Geophys. Res., 106*, 25,759, 2001.

Kanekal, S.G., D.N. Baker, J.B. Blake, Multisatellite measurements of relativistic electrons: Global coherence, *J. Geophys. Res., 106*, 29,721, 2001.

Kim, H.-J., A.A. Chan, R.A. Wolf, and J. Birn, Can substorm produce relativistic outer belt electrons?, *J. Geophys. Res., 105*, 7721, 2000.

Koons. H.C., and D.J. Gorney, A neural network model of the relativistic electron flux at geosynchronous orbit, *J. Geophys. Res., 96*, 5549, 1991.

Lanzerotti, L.J., G.G. Maclennan, and M. Schulz, Radial diffusion of outer-zone electrons: an empirical approach to third-invariant violation, *J. Geophys. Res., 75*, 5351, 1970.

Lanzerotti, L.J., and C.G. Morgan, ULF geomagnetic power near $L$=4 2. Temporal variation of the radial diffusion coefficient for relativistic electrons, *J. Geophys. Res., 78*, 4600, 1973.

Lanzerotti, L.J., D.C. Webb, and C.W. Arthur, Geomagnetic field fluctuations at geosynchronous orbit 2. Radial diffusion, *J. Geophys. Res., 83*, 3866, 1978.

Lezniak T.W., and J.R. Winckler, Experimental study of magnetospheric motions and the acceleration of energetic electrons during substorms, *J. Geophys. Res., 75*, 7075, 1970.

Li X., D. N. Baker, M. Temerin, T.E. Cayton, E.G.D. Reeves, et al., Multisatellite observation of the outer zone electron variation during the November 3-4, 1993, magnetic storm, *J. Geophys. Res., 102*, 14,123, 1997.

Li, X., M. Temerin, D. N. Baker, G. D. Reeves and D. Larson, Quantitative Prediction of Radiation Belt Electrons at Geostationary Orbit Based on Solar Wind Measurements, *Geopys. Res. Lett., 28*, 1887, 2001.

Liu, W.W., and Rostoker G., Energetic Ring Current Particles Generated By Recurring Substorm Cycles, *J. Geophys. Res., 100*, 21,897, 1995.

Liu, W.W., Rostoker G., Baker D.N., Internal Acceleration Of Relativistic Electrons By Large-Amplitude ULF Pulsations, *J. Geophys. Res., 104*, 17391, 1999.

McAdams, K.L., and G.D. Reeves, Non-adiabatic responce of relativistic radiation belt electrons to GEM magnetic storms, *Geopys. Res. Lett., 28*, 1897, 2001.

McIlwain, C.E., Processes acting upon outer zone electrons, in Radiation Belt: Models and Standards, Geophys. Monogr. Ser., vol. 97, edited by J.F. Lemaire, D. Heynderickx, and D.N. Baker, p. 15, AGU, Washington, DC, 1996.

Nagai, T, "Space weather forecast": prediction of relativistic electron intensity at synchronous orbit, *Geophys. Res. Lett., 15*, 425, 1988.

Nishida, A., Outward diffusion of energetic particles from the jovian radiation belt, *J. Geophys. Res., 81*, 1771, 1976.

O'Brien, B.J., C.D. Laughlin, and D.A. Gurnett, High-latitude geophysical studies with satellite Injun 3, *J. Geophys. Res., 69*, 13, 1964.

Paulikas, G.A., and J.B. Blake, Modulation of trapped energetic electrons at 6.6 $R_e$ by the direction of the interplanetary magnetic field, *Geophys. Res. Lett., 3*, 277, 1976.

Paulikas, G.A. and J.B. Blake, Effects of the solar wind on the magnetospheric dynamics: Energetic electrons at the geosynchronous orbit, in Quantitative Modeling of Magnetospheric Processes, Geophys. Monogr. Ser., vol. 21, edited by W.P. Olson, p. 180, AGU, Washington, D.C., 1979.

Paulikas, G.A., J.B. Blake, S.C. Freden, and S.S. Imamoto, Observations of energetic electrons at synchronous altitude 1. General features and diurnal variations, *J. Geophys. Res., 73*, 4915, 1968.

Parks, G.K., and J.R. Winkler, Acceleration of energetic electrons observed at the synchronous altitude during magnetospheric substorms, *J. Geophys. Res.,* 73, 5786, 1968.

Rostoker, G., Geomagnetic indices, *Rev. Geophys. and Space Phys., 10*, 935, 1972.

Rostoker, G., A quantitative relationship between AE and Kp, *J. Geophys. Res., 96*, 5853, 1991.

Rothwell, P., and C. McIlwain, Magnetic storms and the Van Allen radiation belts – observations from satellite 1958ε (Explorer IV), *J. Geophys. Res., 65*, 799, 1960.

Schulz, M., and L.J. Lanzerotti, Particle diffusion in the radiation belts, Springer-Verlag, New York Heidelberg Berlin, 1974.

Troshichev, O.A., V.G. Andrezen, S. Vennerstrom, and E. Friis-Christensen, Magntic activity in the polar cap – a new index, *Planet. Space Sci., 36*, 1095-1102, 1988.

Tsurutani, B.T., W.D. Gonzalez, A.L.C. Gonzalez, F. Tang, J.K. Arballo, and M. Okada, Interplanetary origin of geomagnetic activity in the declining phase of the solar cycle, *J. Geophys. Res., 100*, 21717-21733, 1995.

Van Allen, J.A., and W.C. Lin, Outer radiation belt and solar proton observations with Explorer VII during March-April 1960, *J. Geophys. Res., 65*, 2998, 1960.

Vennerstrom, S., E. Friis-Christensen, O.A. Troshichev, and V.G. Andresen, Comparison between the polar cap index, *PC*, and the auroral electrojet indices *AE*, *AL*, and *AU*, *J. Geophys. Res., 96*, 101, 1991.

West, H.I., Jr. Buck, R.M. Buck, and G.T. Davidson, The dynamics of energetic electrons in the earth's outer radiation belt during 1968 as observed by the Lawrence Livermore National Laboratory's spectrometer on OGO 5, *J. Geophys. Res., 86*, 2111, 1981.

Wilkox, J.M., and N.F. Ness, Quasi-stationary corotating structure in the interplanetary medium, *J. Geophys. Res., 70*, 1815, 1965.

Williams, D.J., A 27-day periodicity in outer zone trapped electron intensities, *J. Geophys. Res., 71*, 1815, 1966.

Williams, D.J., and A.M. Smith, Daytime trapped electron intensities at high latitudes at 1100 kilometers, *J. Geophys. Res., 70*, 541, 1965.

A.V.Dmitriev, Institute of Space Science National Central University, Chung-Li, 320, Taiwan. (e-mail: dalex@jupiter.ss.ncu.edu.tw)

J.-K.Chao, Institute of Space Science National Central University, Chung-Li, 320, Taiwan. (e-mail: jkchao@jupiter.ss.ncu.edu.tw)

Received ________________

**Figure 1.** Time profiles (from top to bottom) of the hourly minimal termination and subsolar magnetopause distances, daily maximal fluxes of >2 MeV electrons observed by GOES-8, *Kp* (multiplied on 10), *Dst* and *PC* indices of geomagnetic activity in 1997 to 2000.

**Figure 2.** Occurrence number distributions for the maximal fluxes of >2 MeV electrons (top panel) and local time of the observations (bottom panel) of peak electron flux (solid histogram) and GREE (dotted histogram).

**Figure 3.** Number of disturbed days per year when the GREEs (solid line) and strong and continuous substorm activity with daily average *Ap*>20 (dashed line) are observed. For comparison with solar activity, the annual sunspot number is shown by dotted line (right axis).

**Figure 4.** Spectra of duration of the relativistic electron enhancements for different years on rising phase of the current 23rd solar cycle.

**Figure 5.** Number of disturbed events per year: the GREE intervals (thick solid line) and strong magnetic storms with *Dst*<-100 nT (thin solid line), *Kp*>6.5 (dashed line) and *PC*>10 (dotted line).

**Figure 6.** Cross-correlation of the GREEs with *Kp* (solid line), *Dst* (dashed dotted line), *PC* (dotted line), *Rs* (gray dotted line) and *Rf* (gray solid line) geomagnetic parameters.

**Figure 7.** Correlation of the GREEs with running averaged values of geomagnetic parameters *Kp* (solid line), *Dst* (dashed dotted line), *PC* (dotted line), *Rs* (gray dotted line) and *Rf* (gray solid line).

**Figure 8.** Scatter plot of (left side) the maximal values of geomagnetic parameters versus 4-day following maximal fluxes in the GREEs and (right panel) daily maximal relativistic electron fluxes versus 4-day preceding daily maximal values of geomagnetic parameters *Kp* (a), *Dst* (b), and *PC* (c).

**Figure 9.** Same as in Fig. 1, but calculated for time interval from April 25 to May 10, 1998.

**Table 1.** The best correlation coefficients $r\,(T_d)$ and $r(T_a)$ between the GREEs and different geomagnetic parameters shifted on $T_d$ hours and averaged on $T_a$ hours respectively.

|       | $T_d$, hours | $N$ | $r\,(T_d)$ | $T_a$, hours | $N$ | $r(T_a)$ |
|-------|------|-----|-----------|------|-----|---------|
| *Dst* | 44   | 263 | -0.3      | 78   | 263 | -0.25   |
| *Kp*  | 45   | 263 | 0.36      | 96   | 263 | 0.38    |
| *PC*  | 3    | 261 | 0.24      | 96   | 262 | 0.17    |
| *Rs*  | 14   | 203 | 0.14      | 182  | 262 | -0.24   |
|       | 75   | 209 | -0.3      |      |     |         |
| *Rf*  | 14   | 203 | 0.17      | -    | -   | -       |
|       | 75   | 209 | -0.28     |      |     |         |

**Table 2.** Correlation coefficients in the multi-parametric linear regression for the GREEs.

| $r_c$ | $Kp\vert_{45}$ | $Dst\vert_{44}$ | $PC\vert_3$ | $Rs\vert_{75}$ | $Rf\vert_{14}$ | $\langle Kp \rangle_{96}$ | $\langle Dst \rangle_{78}$ | $\langle PC \rangle_{96}$ | $\langle Rs \rangle_{182}$ |
|---|---|---|---|---|---|---|---|---|---|
| .60 | .37 | -.36 | .32 | -.34 | .13 | .44 | -.32 | .3 | -.3 |
| .60 | X | -.36 | .32 | -.34 | .13 | .44 | -.32 | .3 | -.3 |
| .58 | .37 | X | .32 | -.34 | .13 | .44 | -.32 | .3 | -.3 |
| .57 | .37 | -.36 | X | -.34 | .13 | .44 | -.32 | .3 | -.3 |
| .60 | .37 | -.36 | .32 | X | .13 | .44 | -.32 | .3 | -.3 |
| .55 | .37 | -.36 | .32 | -.34 | X | .44 | -.32 | .3 | -.3 |
| .57 | .37 | -.36 | .32 | -.34 | .13 | X | -.32 | .3 | -.3 |
| .59 | .37 | -.36 | .32 | -.34 | .13 | .44 | X | .3 | -.3 |
| .60 | .37 | -.36 | .32 | -.34 | .13 | .44 | -.32 | X | -.3 |
| .59 | .37 | -.36 | .32 | -.34 | .13 | .44 | -.32 | .3 | X |
| .60 | X | -.36 | .32 | X | .13 | .44 | -.32 | X | X |
| .57 | X | -.36 | .32 | X | .13 | .44 | -.32 | X | X |
| **.58** | X | -.36 | **.32** | X | **.13** | **.44** | X | X | .3 |
| .56 | X | X | .32 | X | .13 | .44 | X | X | .3 |
| .54 | X | -.36 | X | X | .13 | .44 | X | X | .3 |
| .51 | X | -.36 | .32 | X | X | .44 | X | X | .3 |
| .54 | X | -.36 | .32 | X | .13 | X | X | X | .3 |
| .56 | X | -.36 | .32 | X | .13 | .44 | X | X | X |
| .54 | X | X | .32 | X | .13 | .44 | X | X | X |

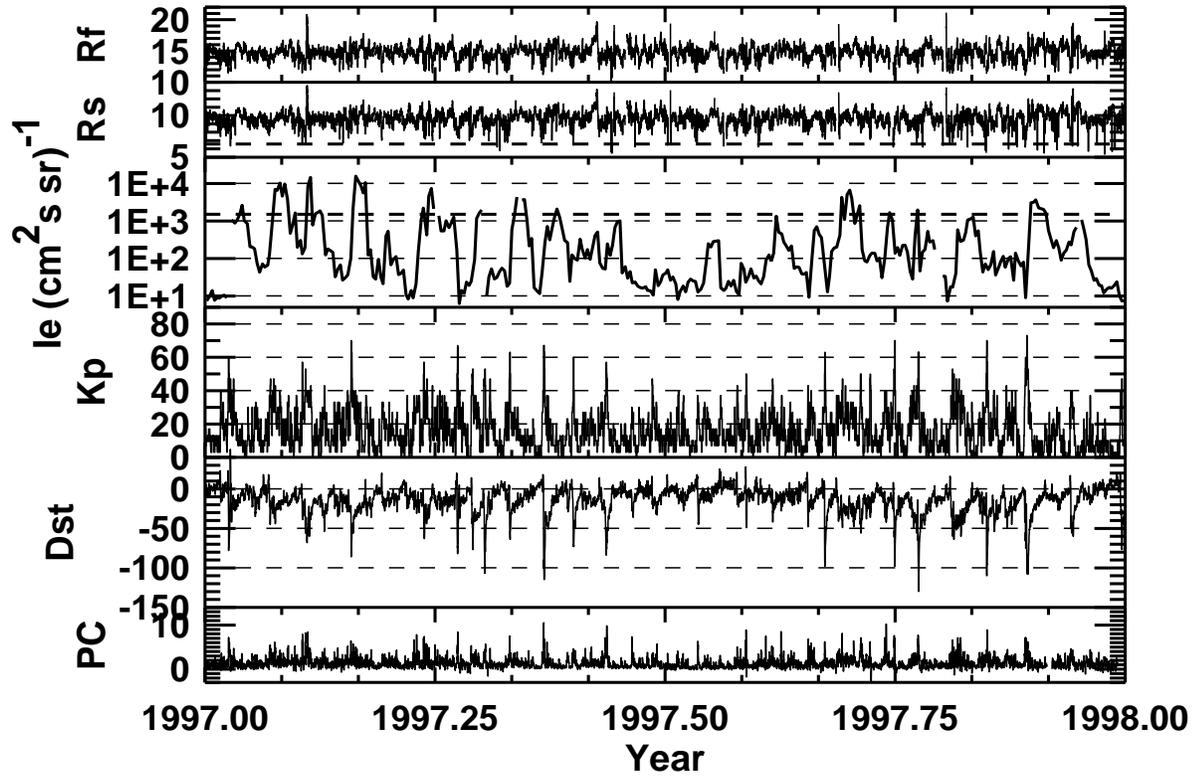

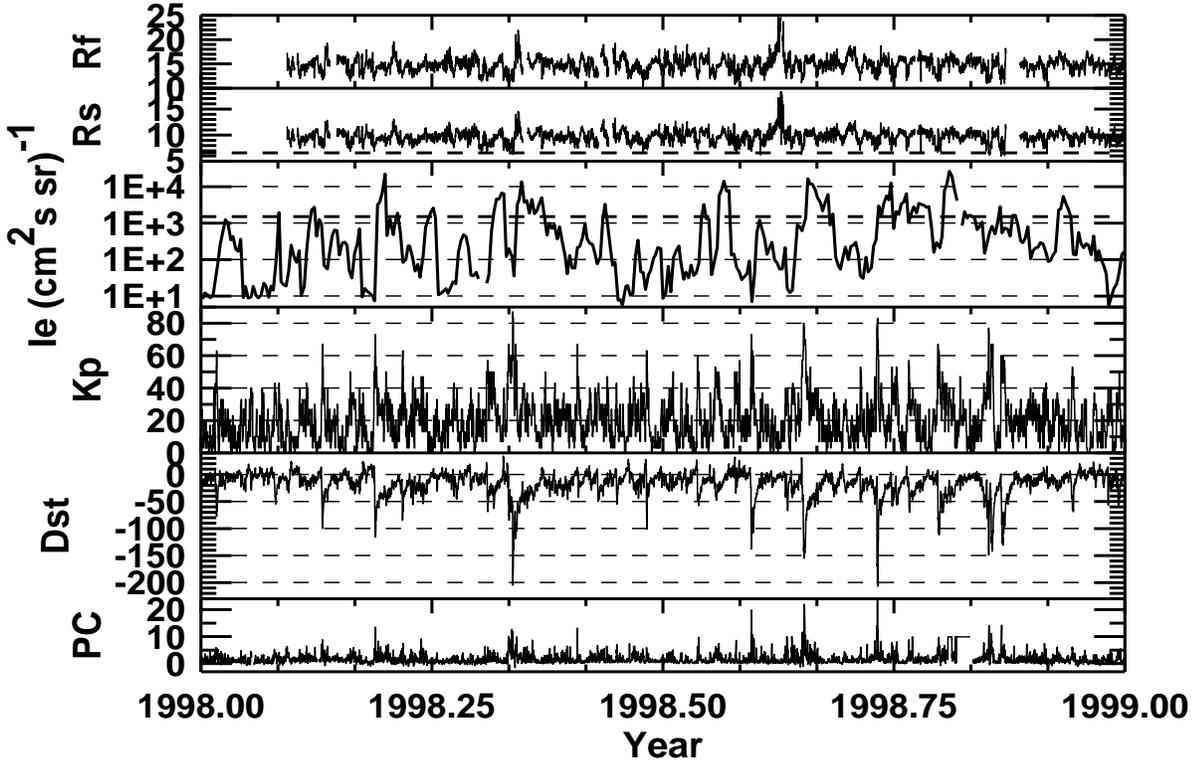

**Figure 1 (beginning)**

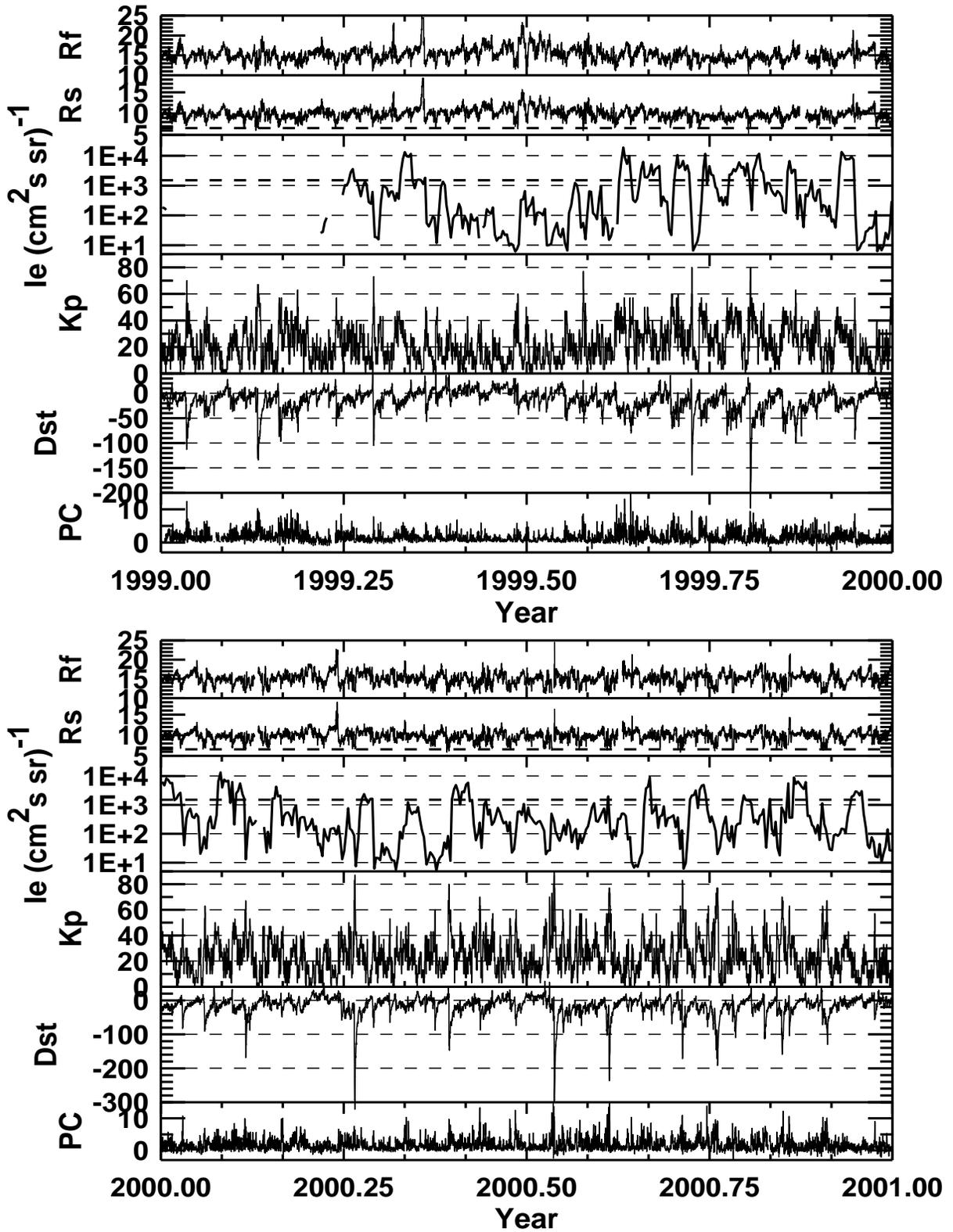

**Figure 1.**

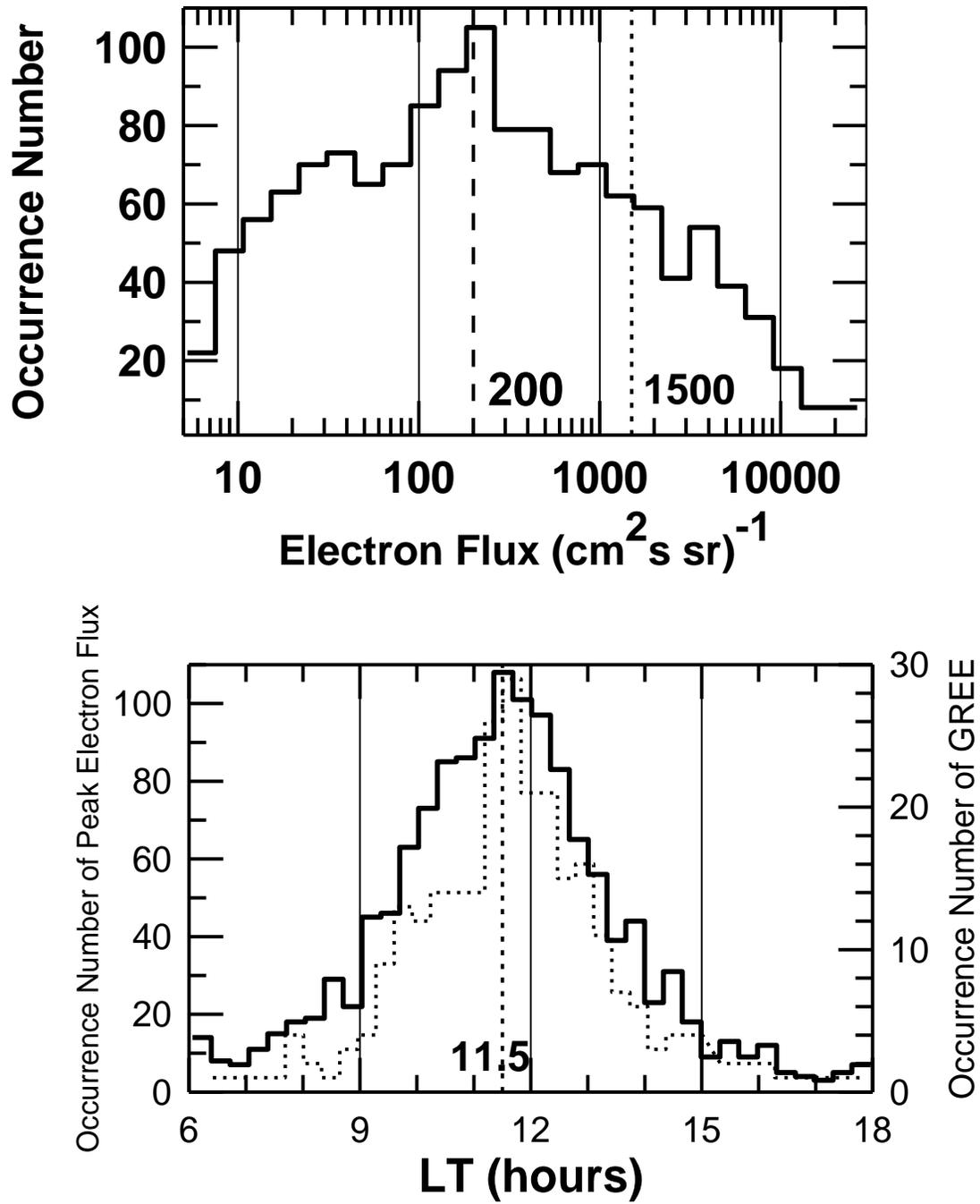

**Figure 2.**

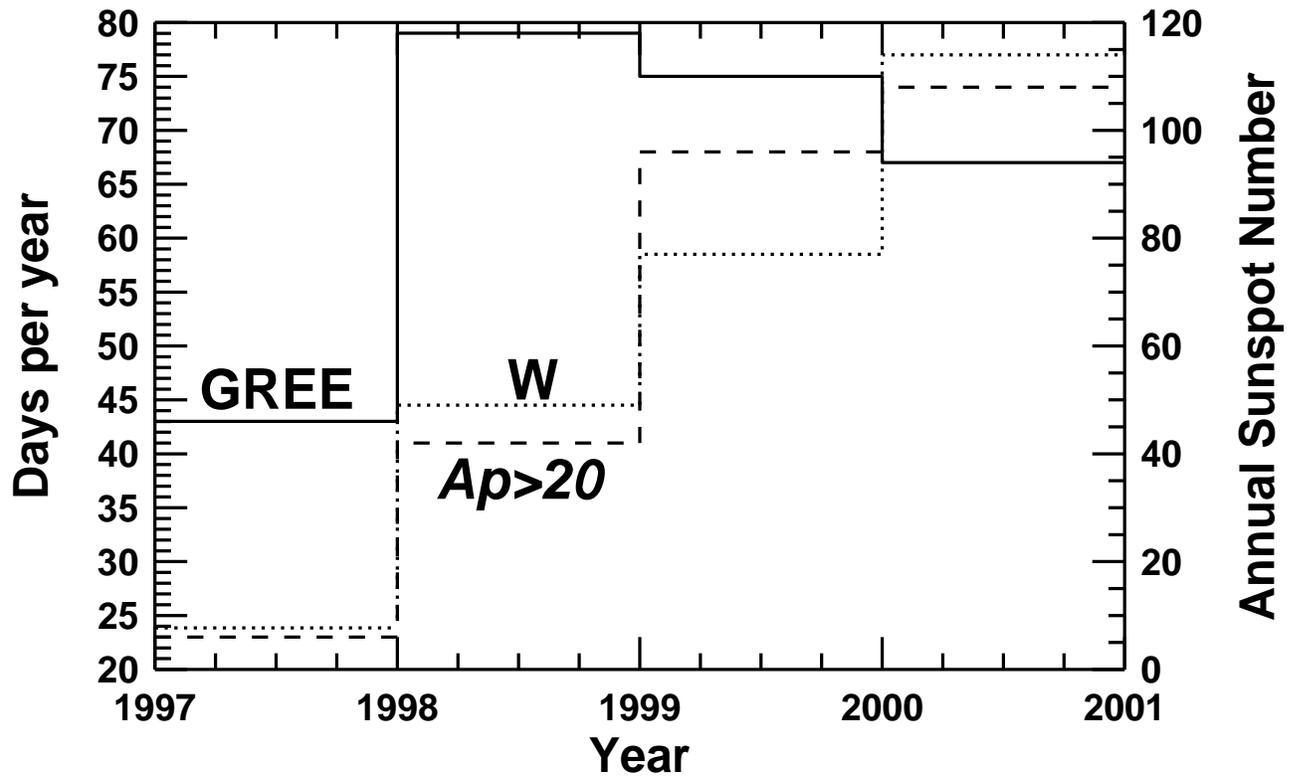

**Figure 3.**

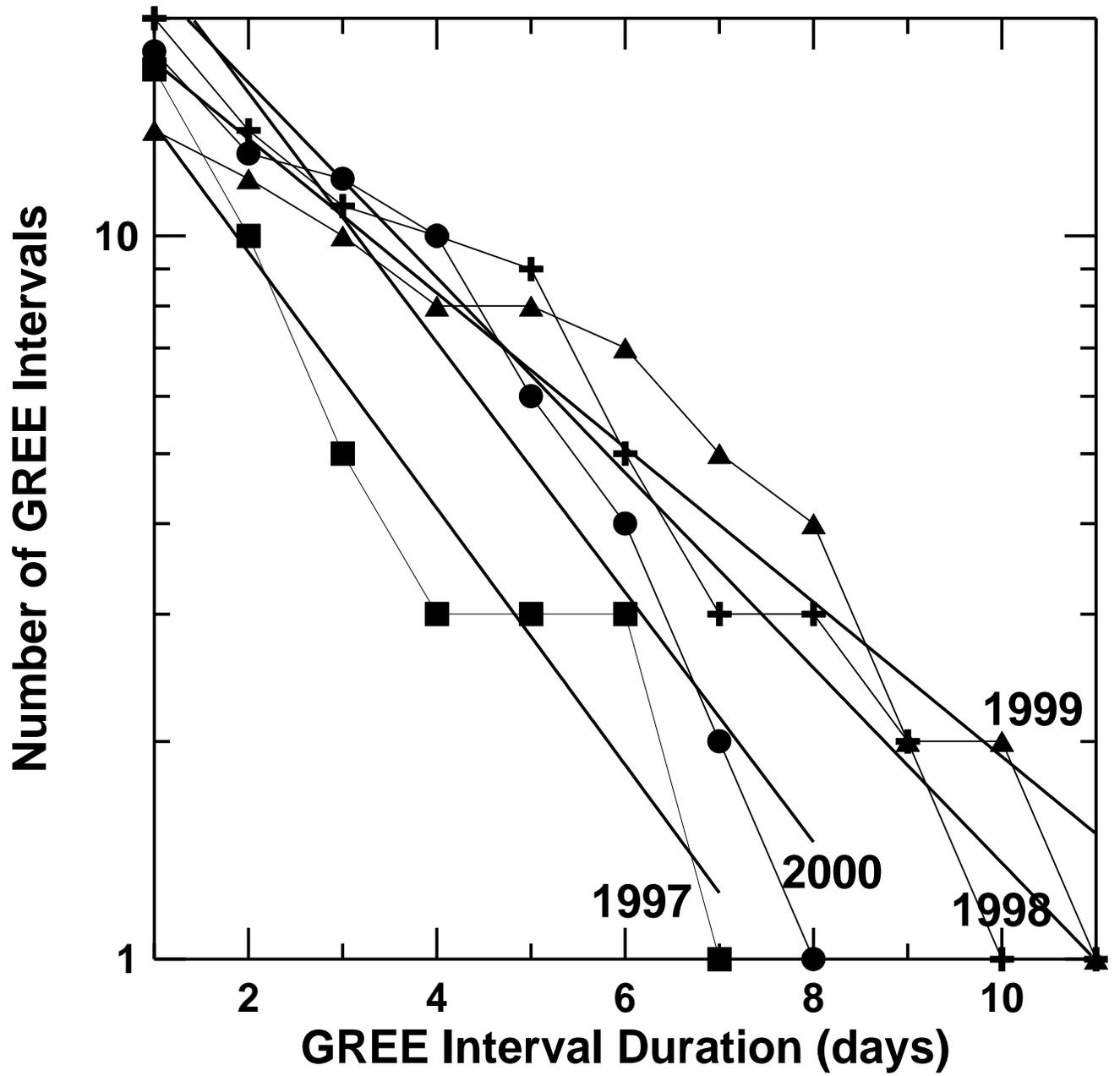

Figure 4.

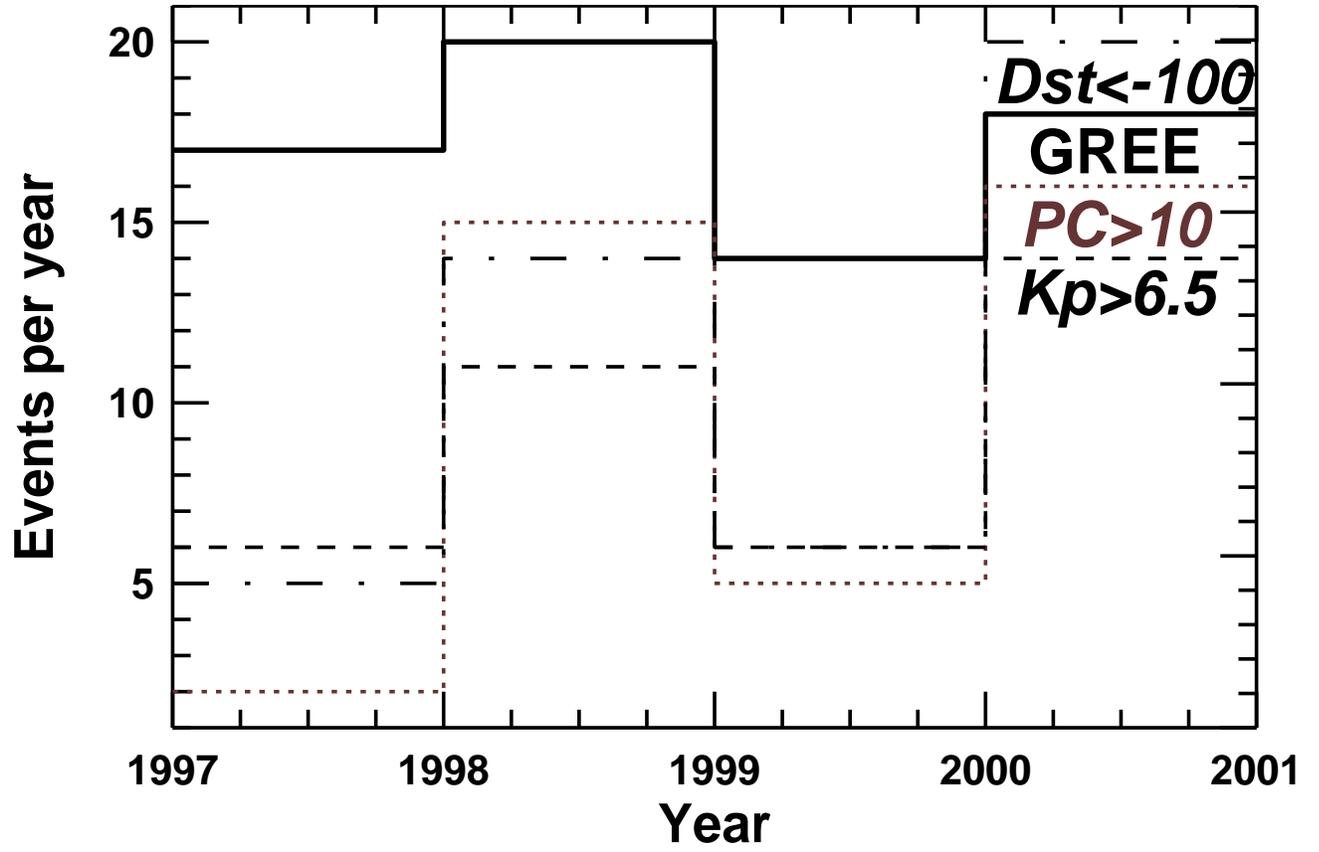

**Figure 5.**

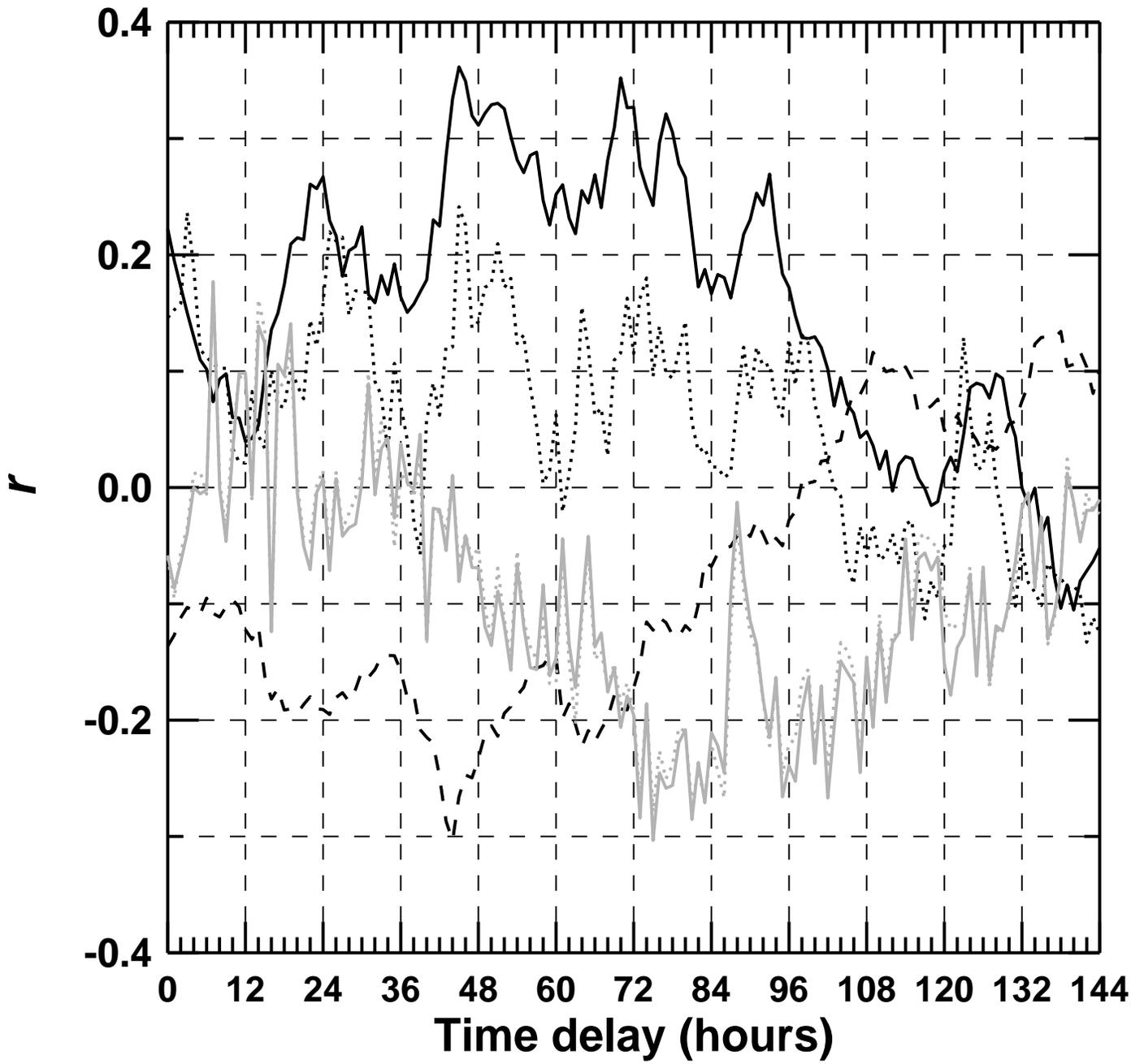

**Figure 6.**

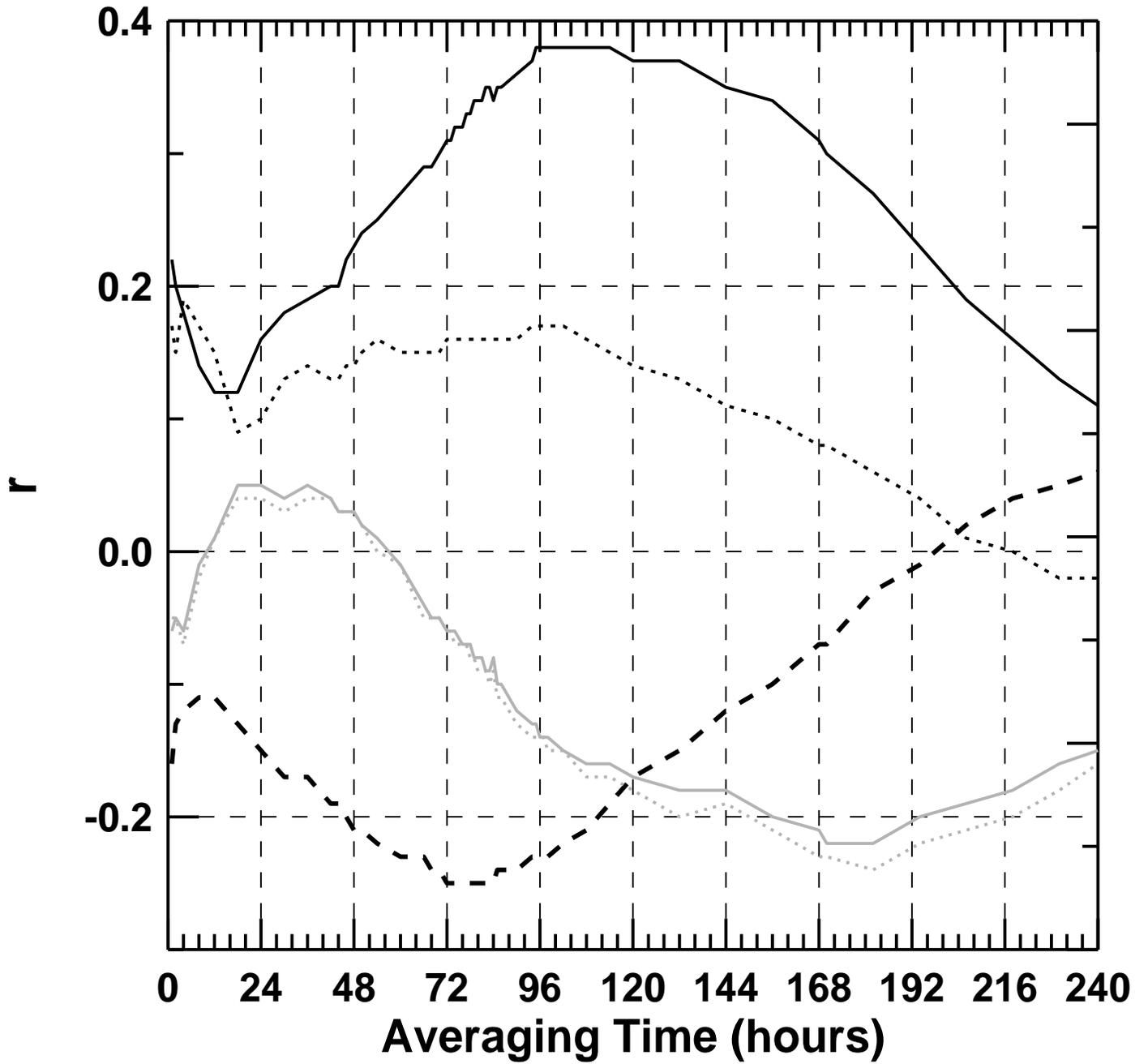

**Figure 7.**

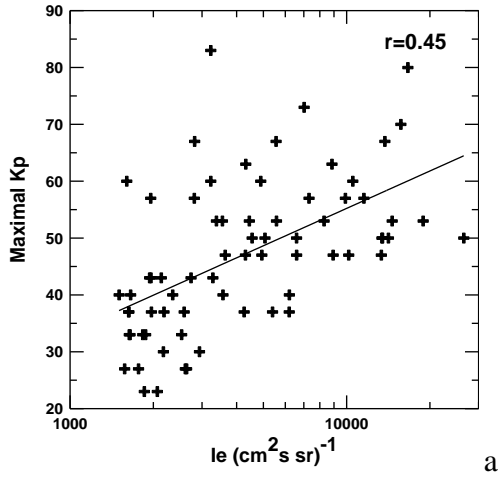

a

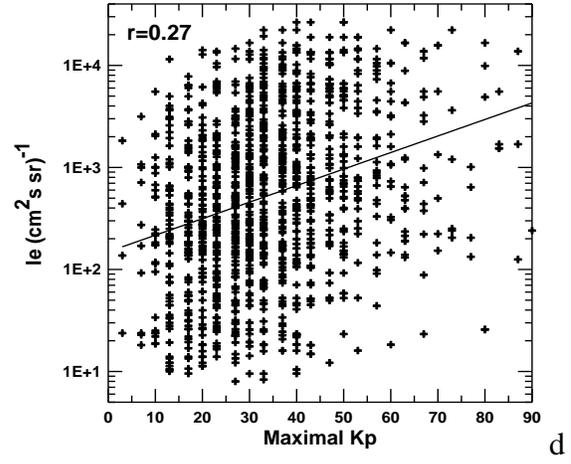

d

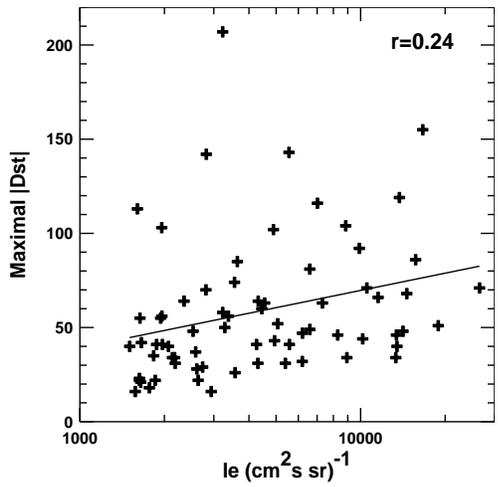

b

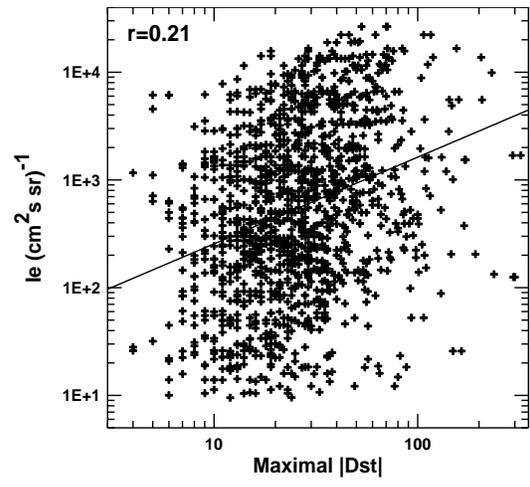

e

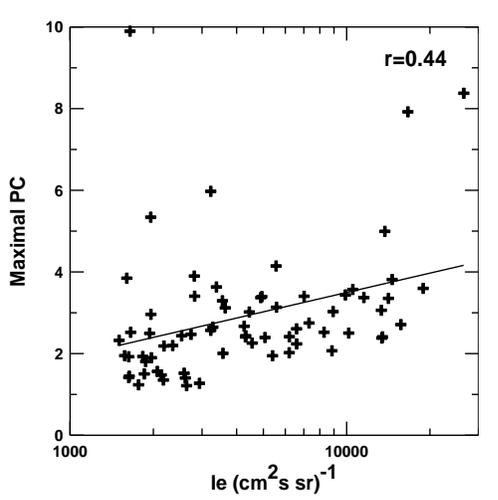

c

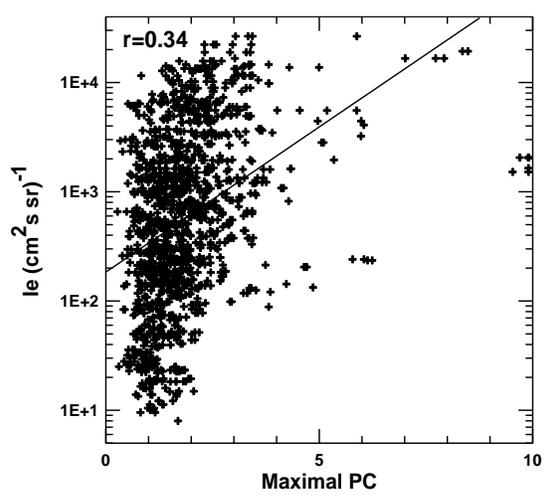

f

**Figure 8.**

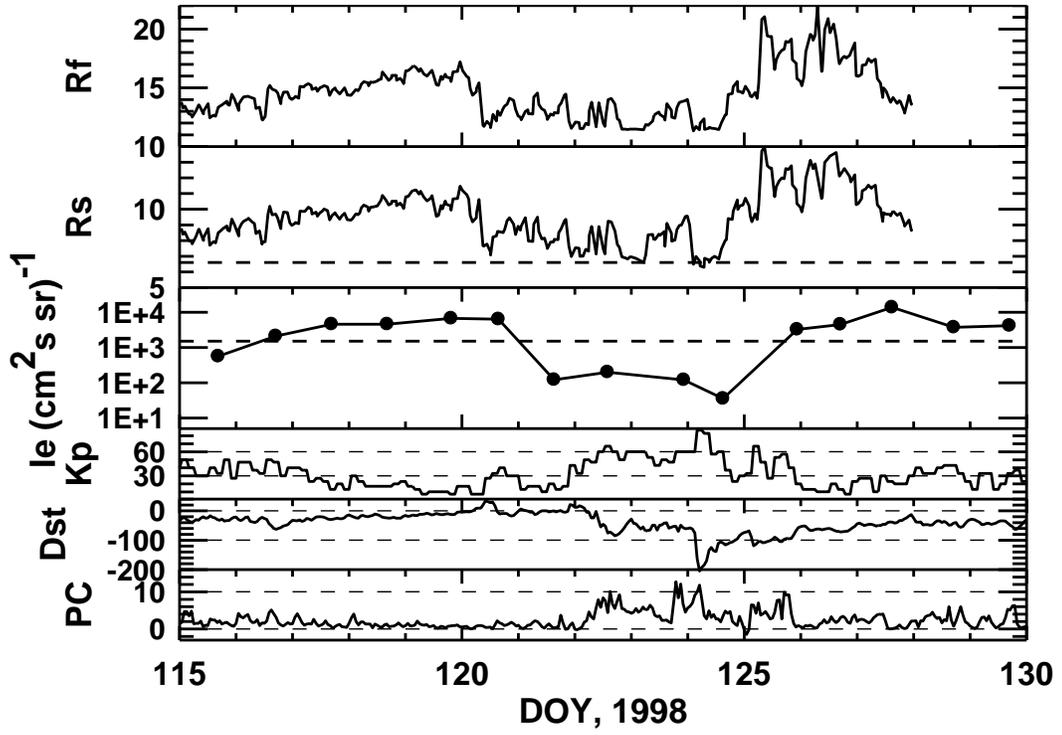